\documentclass[aps,pra,onecolumn,groupedaddress,showpacs,showkeys,superscriptaddress,10pt]{revtex4-1}
\usepackage{graphicx}
\usepackage[colorlinks]{hyperref}
\usepackage{amssymb}
\usepackage{amsmath}
\usepackage{amsfonts}
\usepackage{amssymb}
\usepackage{color}

\begin{document}

\title{Tunable optical second-order sideband effects in a parity-time symmetric optomechanical system}
\author{Xing Xiao}
\affiliation{Department of Electronic Information Engineering, Nanchang University, Nanchang 330031, China}
\author{Qinghong Liao}
\email{nculqh@163.com}
\affiliation{Department of Electronic Information Engineering, Nanchang University, Nanchang 330031, China}
\affiliation{State Key Laboratory of Low-Dimensional Quantum Physics, Department of Physics, Tsinghua University, Beijing 100084, China}

\author{Nanrun Zhou}
\affiliation{Department of Electronic Information Engineering, Nanchang University, Nanchang 330031, China}

\author{Wenjie Nie}
\affiliation{Department of Applied Physics, East China Jiaotong University, Nanchang 330013, China}

\author{Yongchun Liu}
\email{ ycliu@tsinghua.edu.cn}
\affiliation{State Key Laboratory of Low-Dimensional Quantum Physics, Department of Physics, Tsinghua University, Beijing 100084, China}

\begin{abstract}
We theoretically investigate the optical second-order sideband generation (OSSG) in an optical parity-time (PT) symmetric system, which consists of a passive cavity trapping the atomic ensemble and an active cavity. It is found that near the exceptional point (EP), the efficiency of the OSSG increases sharply not only for the blue probe-pump detuning resonant case but also for the red one. Using experimentally achievable parameters, we study the effect of the atomic ensemble on the efficiency of the OSSG. The numerical results show that the efficiency of the OSSG is $30\% $ higher than that of the first-order sideband, which is realized easily by simultaneously modulating the atom-cavity coupling strength and detuning. Moreover, the efficiency of the OSSG can also be tuned effectively by the pump power, and the efficiency is robust when the pump power is strong enough. This study may have some guidance for modulating the nonlinear optical properties and controlling light propagation, which may stimulate further applications in optical communications.

\end{abstract}

\maketitle
\section{INTRODUCTION}
A typical optomechanical system comprises a Fabry-P\'{e}rot cavity~\cite{vitali2007optomechanical} and mechanical oscillator. Cavity optomechanics (COM) describes the interaction between light and the movable mechanical oscillator via the radiation pressure of the system~\cite{kippenberg2007cavity,verhagen2012quantum,liu2016method}, which has become a rapidly developing research field in recent years. Cavity optomechanical systems have potential applications in many fields, such as gravitational wave detectors~\cite{abramovici1992ligo}, photon blockade~\cite{rabl2011photon,huang2018nonreciprocal}, precision measurement~\cite{krause2012high,liu2018room}, force sensors~\cite{wang2015precision}. Moreover, coherent manipulation of light~\cite{lei2015three,zhang2018double}, the optomechanically induced transparency (OMIT)~\cite{nie2017coupling,hou2015optomechanically}, and electromagnetically induced transparency (EIT)~\cite{guo2014electromagnetically}, and high-order sidebands~\cite{xiong2012higher,xiong2016optomechanically,xiong2013carrier,li2016giant,li2018enhanced,he2019parity,chen2019atom} have been studied in detail. The first-order or even high-order sidebands of the pump field can be generated in a cavity optomechanical system, due to the nonlinear optomechanical interactions. Optical high-order harmonic generation was first successfully observed by Franken et al.~\cite{franken1961generation} and has developed potential applications in quantum information processing and optical communication. Nowadays, the second-order sideband~\cite{li2018enhanced,he2019parity,chen2019atom} and the high-order sideband~\cite{xiong2013carrier,yang2017enhanced} have attracted attention for their significant potential applications. Optical high-order sideband generation is related to its input light, and these nonlinear optical effects can only be observed under the condition of strong driving field or special driving pulse. In other words, without certain enhancement of input light, this small nonlinearity is difficult to observe and is detrimental to its application. Cavity optomechanical systems have become an effective way to enhance nonlinear optics, owing to providing extremely high-quality factors with small mode volumes~\cite{louyer2005tunable,vahala2003optical}. However, the second-order and high-order sidebands generated in a cavity optomechanical system are usually much smaller than the first-order sidebands. Therefore, how to generate and enlarge the second-order or high-order sidebands is worthy of our attention.

PT-symmetric systems are open physical systems having balanced loss and gain. Remarkably, such systems can exhibit a phase transition~\cite{bender1998real,bender2007making,mostafazadeh2002pseudo} (spontaneous PT-symmetry breaking). The spontaneous PT-symmetry breaking point is called an exceptional point (EP). PT symmetry has been extensively studied both theoretically~\cite{lin2011unidirectional,jones2012analytic,benisty2011implementation} and experimentally~\cite{ruter2010observation,bittner2012p,regensburger2012parity,peng2014parity}. PT symmetry has important application in various physical systems such as in nonreciprocal light propagation~\cite{peng2014parity,feng2011nonreciprocal}, unidirectional reflectionless resonances, PT-symmetry-breaking chaos~\cite{lu2015p}, and PT-symmetric phonon laser~\cite{jing2014pt}. Recently, some researches~\cite{peng2014parity,li2015enhanced} have shown that PT symmetry can affect the nonlinearity of the system. Peng et al.~\cite{peng2014parity} have successfully observed non-reciprocity in the PT-symmetry-breaking phase due to strong field localization, which significantly enhances nonlinearity. High-order sidebands are the product of optical nonlinearity. Especially in the vicinity of the EP, the transmittance and the second-order sideband effect of the system will be greatly enhanced~\cite{li2016giant,he2019parity,li2015enhanced}. Compared with double-passive optomechanical system, PT-symmetric system can obtain strong effective optical nonlinearity. Consequently, the second-order sideband can be greatly enhanced in the PT-symmetric systems.

On the other hand, the optical cavity with nonlinear medium is driven by a strong pump field and a weak probe field. It can excite multiwave-mixing processes through the optical nonlinear effect~\cite{boyd2003nonlinear}, thus producing a series of optical high-order sidebands. Li et al.~\cite{li2016giant} have used a device of optical coupled microcavities with weak Kerr-type nanostructure materials, and giant enhancement of optical high-order sideband generation and their control were realized. Jiao et al.~\cite{jiao2018optomechanical} have theoretically investigated high-order OMIT process in a nonlinear Kerr resonator, and found that the efficiency of the OSSG could be significantly enhanced by tuning the Kerr coefficient. In addition, in order to enhance the nonlinearity of the system, atoms can be added in the cavity. High-order sidebands and chaos can be observed~\cite{jiang2016optical}. Recently, second-order sideband generation is studied in an optomechanical system with atom-cavity-resonator coupling in Ref.~\cite{chen2019atom}. It is demonstrated that the efficiency of the second-order sideband generation can be significantly enhanced under certain conditions.

Motivated by these developments, we present an optical PT-symmetric system for optical sideband generation. The system is composed of a passive cavity trapping the atomic ensemble and an active cavity, which combines the nonlinearity of atoms and PT symmetry. Therefore, the enhancement of the second-order sideband in this system is predictable. Compared with the traditional optomechanical system, our system possesses some obvious advantages in enhancing the nonlinearity. First, the atom is coupled to the cavity field via the Jaynes-Cummings interaction. The coupling is susceptible to their parameter variations, which enables us to control the high-order sideband of the system flexibly. Second, PT symmetry can be used to enhance the second-order sideband. The relationship between the transmission intensity of the probe field and the oscillating frequency of the mechanical resonator is studied. We also have shown the influence of various factors on the efficiency of the OSSG, including the photon tunneling strength, the atom-cavity coupling strength and detuning, as well as the pump power. Under the modulation of the atom-cavity coupling strength and detuning, we show that the second-order sideband generation is not only with over $150\% $ efficiency, but also $30\% $ higher than that of the first-order sideband generation.

The organization of this paper is as follows. In Sec. II, the theoretical model is introduced and the Hamiltonian of a nonlinear PT-symmetric system is described. The coefficients of the first- and second-order sidebands are deduced. In Sec. III, The relationship between the transmission intensity of the probe field and the oscillating frequency of the mechanical resonator is analyzed. We study the enhancement of the second-order sideband generation in the PT-symmetric system. We also discuss the influence of atom-cavity coupling strength and detuning on the generation of first-order and second-order sidebands. Further, the effect of pump power is also considered. The summary is given in Sec. IV.

\section{THEORETICAL MODEL AND DERIVATION OF HIGH-ORDER SIDEBANDS}

\begin{figure}[htbp]
\centering\includegraphics[width=14cm]{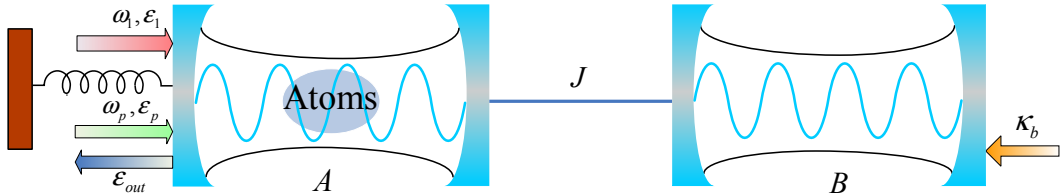}
\caption {Schematic of the hybrid optomechanical system, which consists of a passive cavity and an active cavity. The cavity $A$ traps the atomic ensemble. The two cavities are coupled directly with the photon tunneling strength $J$. The hybrid system is driven by a strong pump field and a weak probe field.}
\label{fig1}
\end{figure}

As shown in Fig.~\ref{fig1}, our hybrid optomechanical system is composed of two cavities (cavity $A$ and cavity $B$). The cavity $A$ is passive with the loss rate ${\kappa _a}$ and couples the mechanical resonator (with resonance frequency ${\omega _m}$, mass $m$ and damping rate ${\Gamma _m}$). The cavity $A$ traps the atomic ensemble with $N$ identical two-level atoms (with transition frequency ${\omega _e}$ and decay rate ${\gamma _a}$). The cavity $B$ is a gain cavity with gain rate ${\kappa _b}$ and couples the cavity $A$ by the coherent photon tunneling strength $J$. The cavity $A$ is driven by a strong pump field with amplitude ${\varepsilon _1}$ with frequency ${\omega _1}$ and a weak probe field ${\varepsilon _p}$ with frequency ${\omega _p}$. The total Hamiltonian of the system in the rotating frame at the frequency ${\omega _1}$ can be written as the following form~\cite{li2018enhanced,liu2018quadrature}:
\begin{eqnarray}\label{1}
H&=&- \hbar {\Delta _1}\left( {{a^\dag }a + {b^\dag }b} \right) + \frac{{{p^2}}}{{2m}} + \frac{1}{2}m\omega _m^2{x^2} + {\Delta _2}{c^\dag }c + \hbar {g_1}x{a^\dag }a
\nonumber\\
 & & {}+ \hbar J\left( {{a^\dag }b + a{b^\dag }} \right) + G\left( {{a^\dag }c + a{c^\dag }} \right)
\nonumber\\
& & {} + i\hbar \sqrt {\eta {\kappa _a}} {\varepsilon _1}\left( {{a^\dag } - a} \right) + i\hbar \sqrt {\eta {\kappa _a}} \left( {{a^\dag }{e^{ - i\Omega t}}{\varepsilon _p} - a{e^{i\Omega t}}\varepsilon _p^ * } \right),
\end{eqnarray}
where $a$ ($b$) and ${a^\dag }$ (${b^\dag }$) denote the annihilation and creation operators of the cavity $A$ ($B$) with resonance frequency ${\omega _a}$ (${\omega _b}$), respectively. For the PT-symmetric structure, we assume that two cavities have the same resonance frequency (${\omega _a} = {\omega _b} = \omega $). $\sigma _{eg}^i = \left| {{g_i}} \right\rangle \left\langle {{e_i}} \right|$ stands for the descending operator of the emitter, with $\left| g \right\rangle$  ($\left\langle e \right|$) being the ground (excited) state. Defining the bosonic annihilation operator for the two-level atomic ensemble $c = \frac{1}{{\sqrt N }}\sum\nolimits_{i = 1}^N {\sigma _{ge}^i} $~\cite{ian2008cavity,wang2016steady}, the interaction between atomic ensemble and cavity $A$ is given as ${H_{ac}} = \sum\nolimits_{i = 1}^N {g\left( {{c_1}\sigma _{eg}^i + c_1^\dag \sigma _{ge}^i} \right)} $. The atom-cavity coupling strength is $G = g\sqrt N $. $p$ ($x$) is the momentum (position) operator of the mechanical resonator with commutation relation $\left[ {x,p} \right] = i\hbar $. The mechanical resonator interacts with the cavity $A$ with optomechanical coupling strength ${g_1} = \left( {{\omega _a}/L} \right)\sqrt {\hbar /m{\omega _m}} $. Here $L$ is the length of the cavity and $m$ is the mass of the mechanical resonator. The frequency detunings between the cavity field, atoms, the pump laser, and the probe laser are depicted by ${\Delta _1} = {\omega _1} - \omega ,{\rm{ }}{\Delta _2} = {\omega _e} - {\omega _1},{\rm{ }}\Omega  = {\omega _p} - {\omega _1},$ respectively. The amplitudes of pump and probe fields are ${\varepsilon _1} = \sqrt {{P_1}/\hbar {\omega _1}} $ and ${\varepsilon _p} = \sqrt {{P_p}/\hbar {\omega _p}} $, respectively. ${P_1}$ (${P_p}$) is the power of a pump (probe) field. $\eta $ is chosen to be the critical coupling ${1 \mathord{\left/
 {\vphantom {1 2}} \right.
 \kern-\nulldelimiterspace} 2}$.

Based on the Hamiltonian in Eq. (\ref{1}), the quantum dynamics of the system can be described by the following Heisenberg-Langevin equation:
\begin{eqnarray}\label{2}
\dot a = \left( {i{\Delta _1} - i{g_1}x - \frac{{{\kappa _a}}}{2}} \right)a - iJb - iGc + \sqrt {\eta {\kappa _a}} {\varepsilon _1} + \sqrt {\eta {\kappa _a}} {\varepsilon _p}{e^{ - i\Omega t}},
\end{eqnarray}
\begin{eqnarray}\label{3}
\dot b = \left( {i{\Delta _1} - \frac{{{\kappa _b}}}{2}} \right)b - iJa,
\end{eqnarray}
\begin{eqnarray}\label{4}
\dot c =  - {\gamma _a}c - i{\Delta _2}c - iGa,
\end{eqnarray}
\begin{eqnarray}\label{5}
\dot x = \frac{p}{m},
\end{eqnarray}
\begin{eqnarray}\label{6}
\dot p =  - m\omega _m^2x - \hbar {g_1}{a^\dag }a - {\Gamma _m}p.
\end{eqnarray}
Considering the perturbation of the probe field, we can regard each operator as the sum of its steady-state value and quantum fluctuation, i.e., $o = {o_s} + \delta o{\rm{ }}$ ($o = a,{\rm{ }}{a^\dag },{\rm{ }}b,{\rm{ }}{b^\dag },{\rm{ }}c,{\rm{ }}{c^\dag },{\rm{ }}x,{\rm{ }}p$), where ${o_s}$ ($\delta o$) are the steady-state solutions (the perturbation terms) of the probe field. The steady-state solutions are obtained as
\begin{eqnarray}\label{7}
{b_s} = \frac{{iJ{a_s}}}{{i{\Delta _1} - \frac{{{\kappa _b}}}{2}}},
\end{eqnarray}
\begin{eqnarray}\label{8}
{c_s} = \frac{{ - iG{a_s}}}{{{\gamma _a} + i{\Delta _2}}},
\end{eqnarray}
\begin{eqnarray}\label{9}
{x_s} = \frac{{ - \hbar {g_1}{{\left| {{a_s}} \right|}^2}}}{{m\omega _m^2}},
\end{eqnarray}
\begin{eqnarray}\label{10}
{a_s} = \frac{{ - \sqrt {\eta {\kappa _a}} {\varepsilon _1}}}{{i\left( {{\Delta _1} - {g_1}{x_s}} \right) - \frac{{{\kappa _a}}}{2} + \frac{{{J^2}}}{{i{\Delta _1} - \frac{{{\kappa _b}}}{2}}} - \frac{{{G^2}}}{{{\gamma _a} + i{\Delta _2}}}}}.
\end{eqnarray}
Further, we consider the optical second-order sideband, but ignore the high-order sideband. We assume that the perturbation terms of Eqs. (\ref{2})-(\ref{6}) have the following forms~\cite{weis2010optomechanically}:
\begin{eqnarray}\label{11}
\delta a = A_1^ - {e^{ - i\Omega t}} + A_1^ + {e^{i\Omega t}} + A_2^ - {e^{ - 2i\Omega t}} + A_2^ + {e^{2i\Omega t}},
\end{eqnarray}
\begin{eqnarray}\label{12}
\delta {a^ * } = {\left( {A_1^ + } \right)^ * }{e^{ - i\Omega t}} + {\left( {A_1^ - } \right)^ * }{e^{i\Omega t}} + {\left( {A_2^ + } \right)^ * }{e^{ - 2i\Omega t}} + {\left( {A_2^ - } \right)^ * }{e^{2i\Omega t}},
\end{eqnarray}
\begin{eqnarray}\label{13}
\delta b = B_1^ - {e^{ - i\Omega t}} + B_1^ + {e^{i\Omega t}} + B_2^ - {e^{ - 2i\Omega t}} + B_2^ + {e^{2i\Omega t}},
\end{eqnarray}
\begin{eqnarray}\label{14}
\delta {b^ * } = {\left( {B_1^ + } \right)^ * }{e^{ - i\Omega t}} + {\left( {B_1^ - } \right)^ * }{e^{i\Omega t}} + {\left( {B_2^ + } \right)^ * }{e^{ - 2i\Omega t}} + {\left( {B_2^ - } \right)^ * }{e^{2i\Omega t}},
\end{eqnarray}
\begin{eqnarray}\label{15}
\delta c = C_1^ - {e^{ - i\Omega t}} + C_1^ + {e^{i\Omega t}} + C_2^ - {e^{ - 2i\Omega t}} + C_2^ + {e^{2i\Omega t}},
\end{eqnarray}
\begin{eqnarray}\label{16}
\delta {c^ * } = {\left( {C_1^ + } \right)^ * }{e^{ - i\Omega t}} + {\left( {C_1^ - } \right)^ * }{e^{i\Omega t}} + {\left( {C_2^ + } \right)^ * }{e^{ - 2i\Omega t}} + {\left( {C_2^ - } \right)^ * }{e^{2i\Omega t}},
\end{eqnarray}
\begin{eqnarray}\label{17}
\delta x = {X_1}{e^{ - i\Omega t}} + X_1^ * {e^{i\Omega t}} + {X_2}{e^{ - 2i\Omega t}} + X_2^ * {e^{2i\Omega t}},
\end{eqnarray}
where the coefficients $A_1^ - $, $A_1^ + $ are the coefficients of first upper and lower sidebands. Similarly, $A_2^ - $, $A_2^ + $ are the coefficients of the second upper and lower sidebands, respectively. We can get these coefficients by substituting Eqs. (\ref{11})-(\ref{17}) into the quantum fluctuation of Eqs. (\ref{2})-(\ref{6}) and compare the coefficients of the same order (the derivation process is in Appendix). The results of the first-order and second-order upper sidebands are obtained, respectively, as
\begin{eqnarray}\label{18}
A_1^ -  = \frac{{\left[ {1 + if\left( \Omega  \right)} \right]\sqrt {\eta {\kappa _a}} {\varepsilon _p}}}{{ - {\beta _2}\left( \Omega  \right)\left[ {1 + if\left( \Omega  \right)} \right] - i\hbar g_1^2{{\left| {{a_s}} \right|}^2}\chi \left( \Omega  \right)}},
\end{eqnarray}
\begin{eqnarray}\label{19}
A_2^ -  = \frac{{i{g_1}\left\{ {{a_s}\left[ {{\theta _1}\left( \Omega  \right) - {\theta _3}\left( \Omega  \right)} \right] + {\theta _4}\left( \Omega  \right)} \right\}}}{{{\beta _2}\left( {2\Omega } \right) + i{g_1}{a_s}{\theta _2}\left( \Omega  \right)}},
\end{eqnarray}
with
$\bar \Delta  = {\Delta _1} - {g_1}{x_s},$ ${\alpha _1}\left( \Omega  \right) = \frac{{{J^2}}}{{\frac{{{\kappa _b}}}{2} + i\left( {{\Delta _1} - \Omega } \right)}},$ ${\alpha _2}\left( \Omega  \right) = \frac{{{G^2}}}{{{\gamma _a} - i\left( {{\Delta _2} + \Omega } \right)}},$
$\chi \left( \Omega  \right) = {1 \mathord{\left/
 {\vphantom {1 {m\left( {\omega _m^2 - {\Omega ^2} - i\Omega {\Gamma _m}} \right)}}} \right.
 \kern-\nulldelimiterspace} {m\left( {\omega _m^2 - {\Omega ^2} - i\Omega {\Gamma _m}} \right)}},$
 ${\beta _1}\left( \Omega  \right) = i\left( {\bar \Delta  - \Omega } \right) + \frac{{{\kappa _a}}}{2} + {\alpha _1}\left( \Omega  \right) + {\alpha _2}\left( \Omega  \right),$
 $f\left( \Omega  \right) = \frac{{\hbar g_1^2{{\left| {{a_s}} \right|}^2}\chi \left( \Omega  \right)}}{{{\beta _1}\left( \Omega  \right)}},$
 ${X_1} = \frac{{ - \hbar {g_1}a_s^ * \chi \left( \Omega  \right)}}{{1 + if\left( \Omega  \right)}}A_1^ - ,$
 $A{_1^ {+ ^ *} } = \frac{{i{g_1}a_s^ * {X_1}}}{{{\beta _1}\left( \Omega  \right)}}$,
 ${\alpha _3}\left( \Omega  \right) = \frac{{{J^2}}}{{\frac{{{\kappa _b}}}{2} - i\left( {{\Delta _1} + \Omega } \right)}},$
 ${\alpha _4}\left( \Omega  \right) = \frac{{{G^2}}}{{{\gamma _a} + i\left( {{\Delta _2} - \Omega } \right)}},$
 ${\beta _2}\left( \Omega  \right) = i\left( {\bar \Delta  + \Omega } \right) - \frac{{{\kappa _a}}}{2} - {\alpha _3}\left( \Omega  \right) - {\alpha _4}\left( \Omega  \right),$
 ${\alpha _1}\left( {2\Omega } \right) = \frac{{{J^2}}}{{\frac{{{\kappa _b}}}{2} + i\left( {{\Delta _1} - 2\Omega } \right)}},$
 ${\alpha _2}\left( {2\Omega } \right) = \frac{{{G^2}}}{{{\gamma _a} - i\left( {{\Delta _2} + 2\Omega } \right)}},$
 ${\beta _1}\left( {2\Omega } \right) = i\left( {\bar \Delta  - 2\Omega } \right) + \frac{{{\kappa _a}}}{2} + {\alpha _1}\left( {2\Omega } \right) + {\alpha _2}\left( {2\Omega } \right),$
 ${\theta _4}\left( \Omega  \right) = A_1^ - {X_1},$
 $\chi \left( {2\Omega } \right) = {1 \mathord{\left/
 {\vphantom {1 {m\left( {\omega _m^2 - 4{\Omega ^2} - 2i\Omega {\Gamma _m}} \right)}}} \right.
 \kern-\nulldelimiterspace} {m\left( {\omega _m^2 - 4{\Omega ^2} - 2i\Omega {\Gamma _m}} \right)}},$
 $f\left( {2\Omega } \right) = \frac{{\hbar g_1^2{{\left| {{a_s}} \right|}^2}\chi \left( {2\Omega } \right)}}{{{\beta _1}\left( {2\Omega } \right)}},$
 ${\theta _1}\left( \Omega  \right) = \frac{{{g_1}X_1^2f\left( {2\Omega } \right)}}{{{\beta _1}\left( \Omega  \right)\left[ {1 + if\left( {2\Omega } \right)} \right]}},$
 ${\theta _2}\left( \Omega  \right) = \frac{{\hbar {g_1}a_s^*\chi \left( {2\Omega } \right)}}{{1 + if\left( {2\Omega } \right)}},$
 ${f_1}\left( \Omega  \right) = \frac{{\hbar g_1^2a_s^*{\theta _4}\left( \Omega  \right)\chi \left( {2\Omega } \right)}}{{{\beta _1}\left( \Omega  \right)}},$
 ${\theta _3}\left( \Omega  \right) = \frac{{i{f_1}\left( \Omega  \right)}}{{1 + if\left( {2\Omega } \right)}},$
 ${\alpha _3}\left( {2\Omega } \right) = \frac{{{J^2}}}{{\frac{{{\kappa _b}}}{2} - i\left( {{\Delta _1} + 2\Omega } \right)}},$
 ${\alpha _4}\left( {2\Omega } \right) = \frac{{{G^2}}}{{{\gamma _a} + i\left( {{\Delta _2} - 2\Omega } \right)}},$
 ${\beta _2}\left( {2\Omega } \right) = i\left( {\bar \Delta  + 2\Omega } \right) - \frac{{{\kappa _a}}}{2} - {\alpha _3}\left( {2\Omega } \right) - {\alpha _4}\left( {2\Omega } \right).$

Further, based on the input-output relation ${S_{out}} = {S_{in}} - \sqrt {\eta {\kappa _a}} a$~\cite{li2018enhanced,weis2010optomechanically}, we can obtain the output field as:
\begin{eqnarray}\label{20}
{S_{out}} = {d_1}{e^{ - i{\omega _1}t}} + {d_p}{e^{ - i{\omega _p}t}} - \sqrt {\eta {\kappa _a}} A_2^ - {e^{ - i\left( {2{\omega _p} - {\omega _1}} \right)t}}
\nonumber\\
 - \sqrt {\eta {\kappa _a}} A_1^ + {e^{ - i\left( {2{\omega _1} - {\omega _p}} \right)t}} - \sqrt {\eta {\kappa _a}} A_2^ + {e^{ - i\left( {3{\omega _1} - 2{\omega _p}} \right)t}},
\end{eqnarray}
where ${d_1} = {\varepsilon _1} - \sqrt {\eta {\kappa _a}} {a_s}, {\rm{ }}{d_p} = {\varepsilon _p} - \sqrt {\eta {\kappa _a}} A_1^ - .$ The terms ${d_1}{e^{ - i{\omega _1}t}}$ and ${d_p}{e^{ - i{\omega _p}t}}$ represent the output fields with the frequencies of ${\omega _1}$ and ${\omega _p}$, respectively. The transmission of the probe field can be defined as ${t_p} = {d_p}/{\varepsilon _p}$, the optical transmission rate can be showed as
\begin{eqnarray}\label{21}
{\left| {{t_p}} \right|^2} = {\left| {1 - \frac{{\sqrt {\eta {\kappa _a}} A_1^ - }}{{{\varepsilon _p}}}} \right|^2}.
\end{eqnarray}
The term $ - \sqrt {\eta {\kappa _a}} A_1^ + {e^{ - i\left( {2{\omega _1} - {\omega _p}} \right)t}}$ describes the Stokes process. The terms $ - \sqrt {\eta {\kappa _a}} A_2^ - {e^{ - i\left( {2{\omega _p} - {\omega _1}} \right)t}}$ and $ - \sqrt {\eta {\kappa _a}} A_2^ + {e^{ - i\left( {3{\omega _1} - 2{\omega _p}} \right)t}}$ represent the output fields with the frequencies of ${\omega _1} + 2\Omega $ and ${\omega _1} - 2\Omega $, respectively. The former is the upper OSSG process and the latter is the lower OSSG process. We give the efficiency of the optical second-order generation (OSSG) as
\begin{eqnarray}\label{22}
{\eta _s} = \left| { - \frac{{\sqrt {\eta {\kappa _a}} A_2^ - }}{{{\varepsilon _p}}}} \right|.
\end{eqnarray}
And the efficiency of the optical first-order sideband generation (OFSG) as
\begin{eqnarray}\label{23}
{\eta _f} = \left| { - \frac{{\sqrt {\eta {\kappa _a}} A_1^ - }}{{{\varepsilon _p}}}} \right|.
\end{eqnarray}

\section{RESULTS AND DISCUSSION}

Based on the hybrid cavity optomechanical system, the properties of the OSSG for the system are studied by using the realizable experimental parameters~\cite{li2018enhanced,chen2019atom,weis2010optomechanically,gigan2006self,arcizet2006radiation,genes2008emergence}: ${\omega _m} = 2\pi  \times 10{\rm{ MHz}}$, $m = 20{\rm{ ng}}$, ${\Gamma _m} = 2\pi  \times 40{\rm{ kHz}}$, $L = 1.0{\rm{ mm}}$, ${\kappa _a} = 2\pi  \times 2{\rm{ MHz}}$, $G = 2\pi  \times 10{\rm{ MHz}}$, ${\gamma _a} = 2\pi  \times 5{\rm{ MHz}}$, $\eta  = 1/2$, $\lambda  = 1064{\rm{ nm}}$, ${P_1} = 310.3{\rm{ }}\mu {\rm{W}}$, ${\varepsilon _p} = 0.05{\varepsilon _1}$, ${\Delta _1} =  - {\omega _m}$, ${\Delta _2} = {\omega _m}$.


\subsection{DEPENDENCE OF THE TRANSMISSION INTENSITY OF THE PROBE FIELD ON THE OSCILLATING FREQUENCY OF THE MECHANICAL RESONATOR}

\begin{figure}[htbp]
\centering\includegraphics[width=14cm]{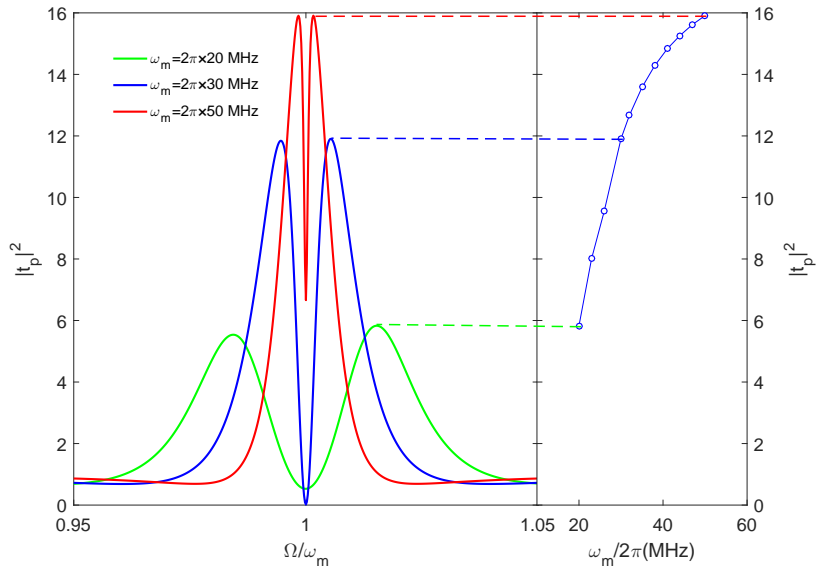}
\caption {Simulation results of the transmission intensity of probe field ${\left| {{t_p}} \right|^2}$ as a function of the detuning $\Omega /{\omega _m}$ for three different values of the oscillating frequency of the mechanical resonator ${\omega _m}$ from $2\pi  \times 20{\rm{ MHz}}$ to $2\pi  \times 50{\rm{ MHz}}$ (corresponding to the curves from bottom to up in the figure). The maximum value of the transmission intensity directly indicates the maximum probe power transmission achieved in this case. These values are given in the right-hand panel for a large scan of ${\omega _m}$, representing the relationship between the transmission intensity of probe field ${\left| {{t_p}} \right|^2}$ and the oscillating frequency of the mechanical resonator ${\omega _m}$. The other parameters are $m = 20{\rm{ ng}}$, ${\Gamma _m} = 2\pi  \times 40{\rm{ kHz}}$, $L = 1.0{\rm{ mm}}$, ${\kappa _a} = 2\pi  \times 2{\rm{ MHz}}$, ${\kappa _b} =  - {\kappa _a}$, $J = 0.45{\kappa _a}$, $G = 0$, ${\gamma _a} = 2\pi  \times 5{\rm{ MHz}}$, $\eta  = 1/2$, $\lambda  = 1064{\rm{ nm}}$, ${P_1} = 310.3{\rm{ }}\mu {\rm{W}}$, ${\varepsilon _p} = 0.05{\varepsilon _1}$, ${\Delta _1} =  - {\omega _m}$, ${\Delta _2} = {\omega _m}$.}
\label{fig2}
\end{figure}

We firstly investigate the influence of the oscillating frequency of the mechanical resonator on the transmission without atoms. Compared with the conventional OMIT profile in a lossy cavity optomechanical system, as can be seen from Fig.~\ref{fig2} that an inverted-OMIT profile~\cite{jing2015optomechanically,oishi2013inverted} of the transmission spectrum occurs when the system has PT-symmetric structure, manifested as an absorbed valley between two symmetric strongly amplifying peaks. The physical interpretation for these phenomena are as follows. With the help of the gain cavity, the absorption valleys caused by the loss cavity were filled and converted into amplifying peaks, indicating that the input field is efficiently amplified and thus the optical transmission intensity ${\left| {{t_p}} \right|^2}$ is amplified. Furthermore, the transmission intensity is far greater than 1. When the oscillating frequency of the mechanical resonator is ${\omega _m} = 20{\rm{ MHz}}$ or ${\omega _m} = 30{\rm{ MHz}}$, ${\left| {{t_p}} \right|^2}$ is very low near $\Omega /{\omega _m} = 1$. This means that the probe field is almost completely absorbed under the resonance condition. With increased values of ${\omega _m}$, the transmission intensity will also become large. When it is greater than $50{\rm{ MHz}}$, the transmission intensity will tend to be stable. The relationship between the transmission intensity of the probe field and the oscillating frequency of the mechanical resonator includes linearity and nonlinearity, as shown in the right-hand panel of Fig.~\ref{fig2}. In linear regime, the coupled double-cavity system with balanced gain and loss is PT-symmetric structure. In nonlinear regime, when the frequency of the mechanical oscillator is greater than $2\pi  \times 30{\rm{ MHz}}$, the gain-saturation nonlinearity leads to strong field localization in the gain cavity. In this case, the gain of the active cavity no longer matches the loss of the passive cavity.

\subsection{DEPENDENCE OF THE SECOND-ORDER SIDEBAND ON THE PHOTON-TUNNELING STRENGTH $J$}
\begin{figure}[htbp]
\centering\includegraphics[width=12cm]{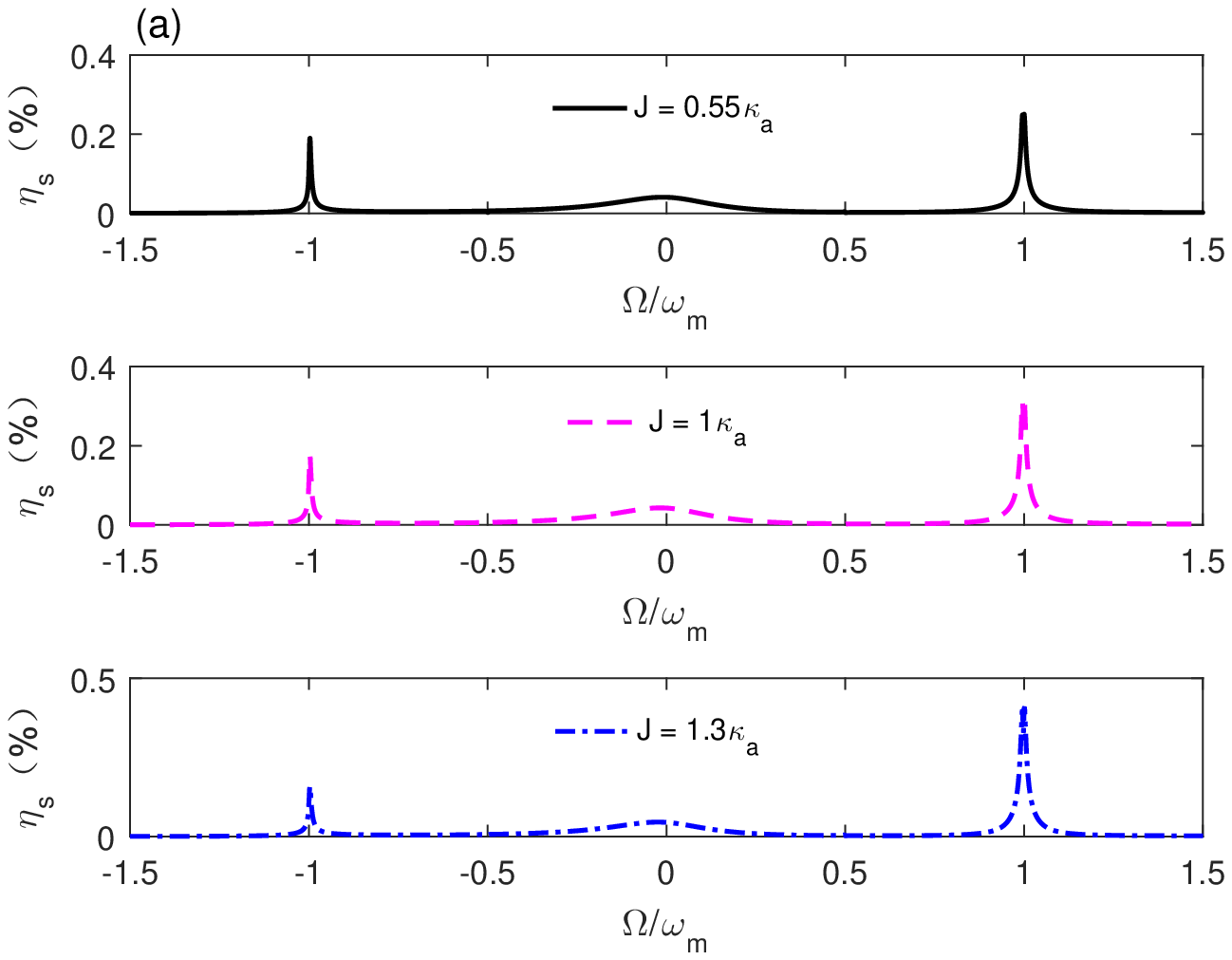}
\includegraphics[width=12cm]{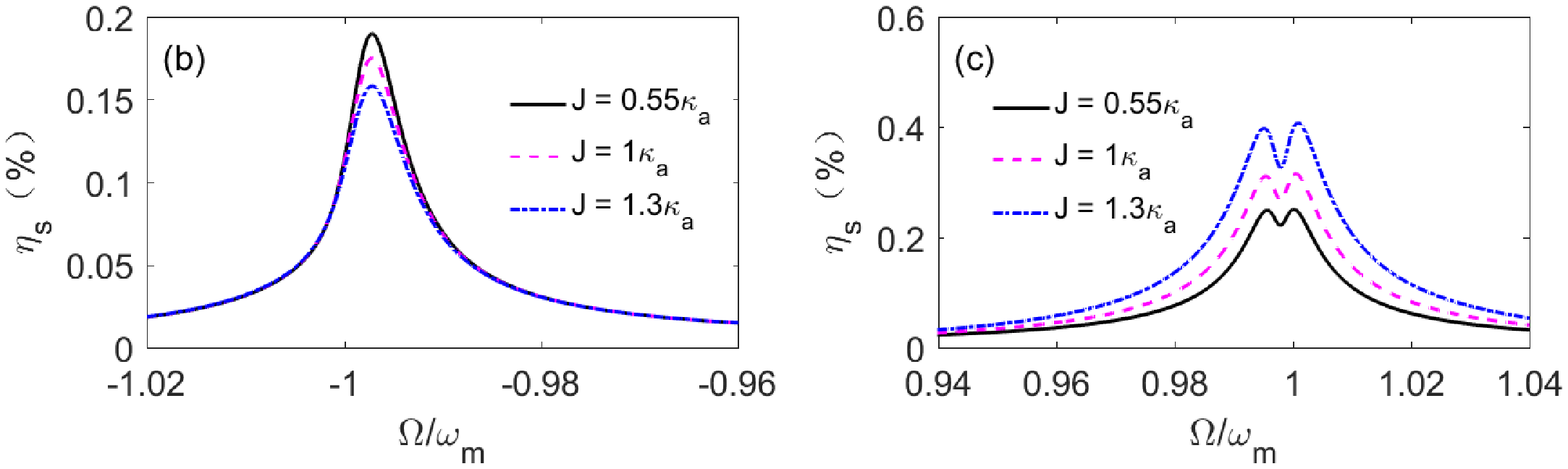}
\caption {(a) Efficiency ${\eta _s}$ of the OSSG as a function of the detuning $\Omega /{\omega _m}$ for different the coupling strengths $J = 0.55{\kappa _a}$, $J = 1{\kappa _a}$, $J = 1.3{\kappa _a}$. (b) Magnified OSSG peak near $\Omega /{\omega _m} =  - 1$. (c) Magnified OSSG peak near $\Omega /{\omega _m} = 1$. The other parameter values are the same as in Fig.~\ref{fig2} except for ${\omega _m} = 2\pi  \times 10{\rm{ MHz,}}$ $G = 2\pi  \times 10{\rm{ MHz}}$.}
\label{fig3}
\end{figure}

The passive cavity trapping the atomic ensemble, for different the coupling strengths $J$ between the loss and gain cavity, we plot the efficiency of the OSSG as a function of the detuning $\Omega /{\omega _m}$ in Fig.~\ref{fig3}(a). According to Fig.~\ref{fig3}(b), we can see clearly that the amplification peak of the OSSG is near $\Omega /{\omega _m} =  - 1$, which is attributed to the coupling between atom and cavity. In the ordinary optomechanical system without atoms, the peak of the OSSG around $\Omega /{\omega _m} =  - 1$ generally will not appear. The first-order sideband generated in an optomechanical system without atoms is completely suppressed. Moreover, with the increase of coupling strength $J$, the peak value decreases obviously. On the contrary, with the increase of coupling strength, the efficiency of the OSSG near $\Omega /{\omega _m} = 1$ increases, as is shown in Fig.~\ref{fig3}(a) and Fig.~\ref{fig3}(c). This shows that the gain compensates the loss in the passive cavity, which leads to the sideband amplification of the transmission spectrum. We note that the condition of reaching the maximum value of ${\eta _s}$ is not exactly the resonance condition $\Omega /{\omega _m} = 1$ but shift a small frequency in the presence.

\begin{figure}[htbp]
\centering\includegraphics[width=7cm]{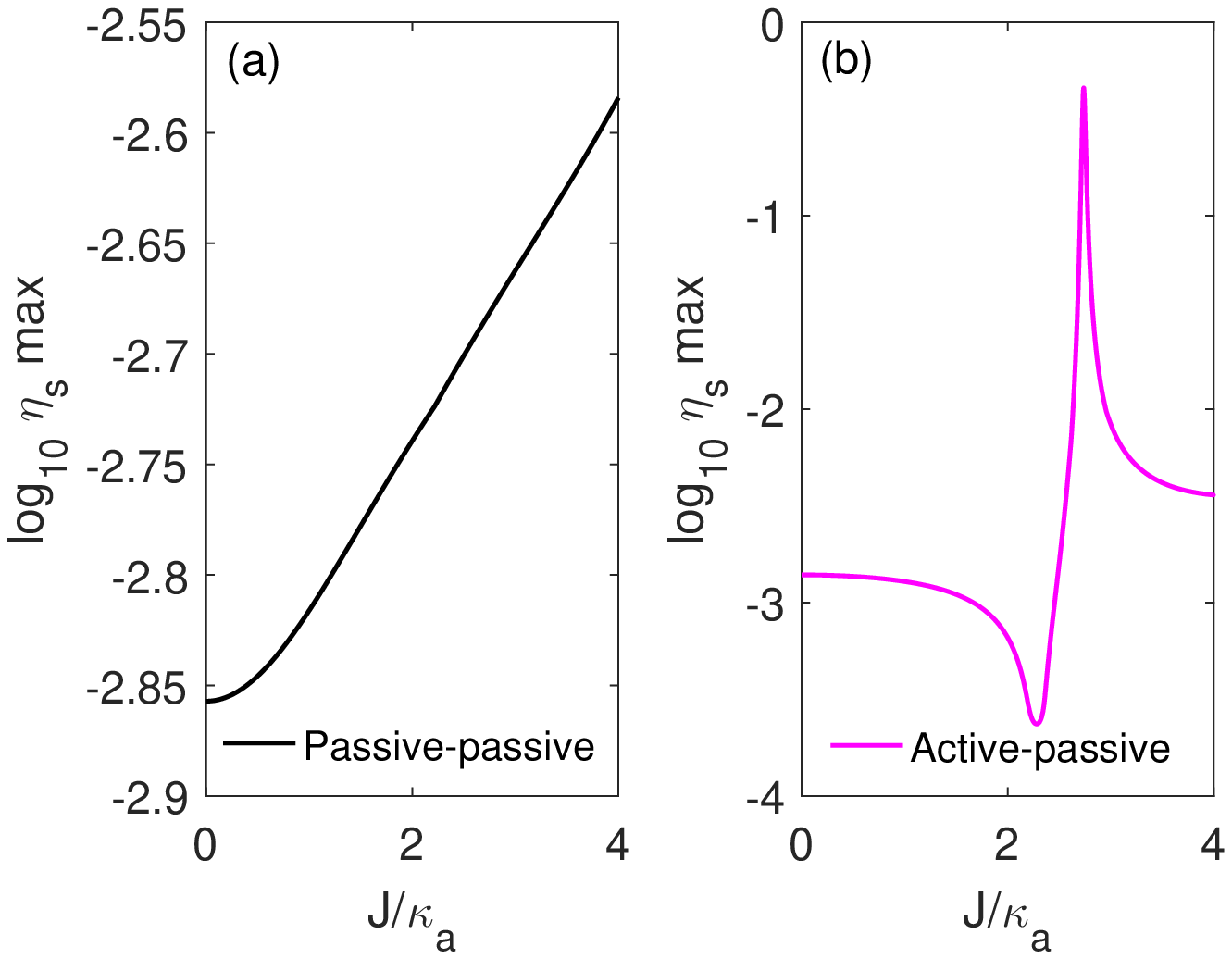}
\includegraphics[width=7cm]{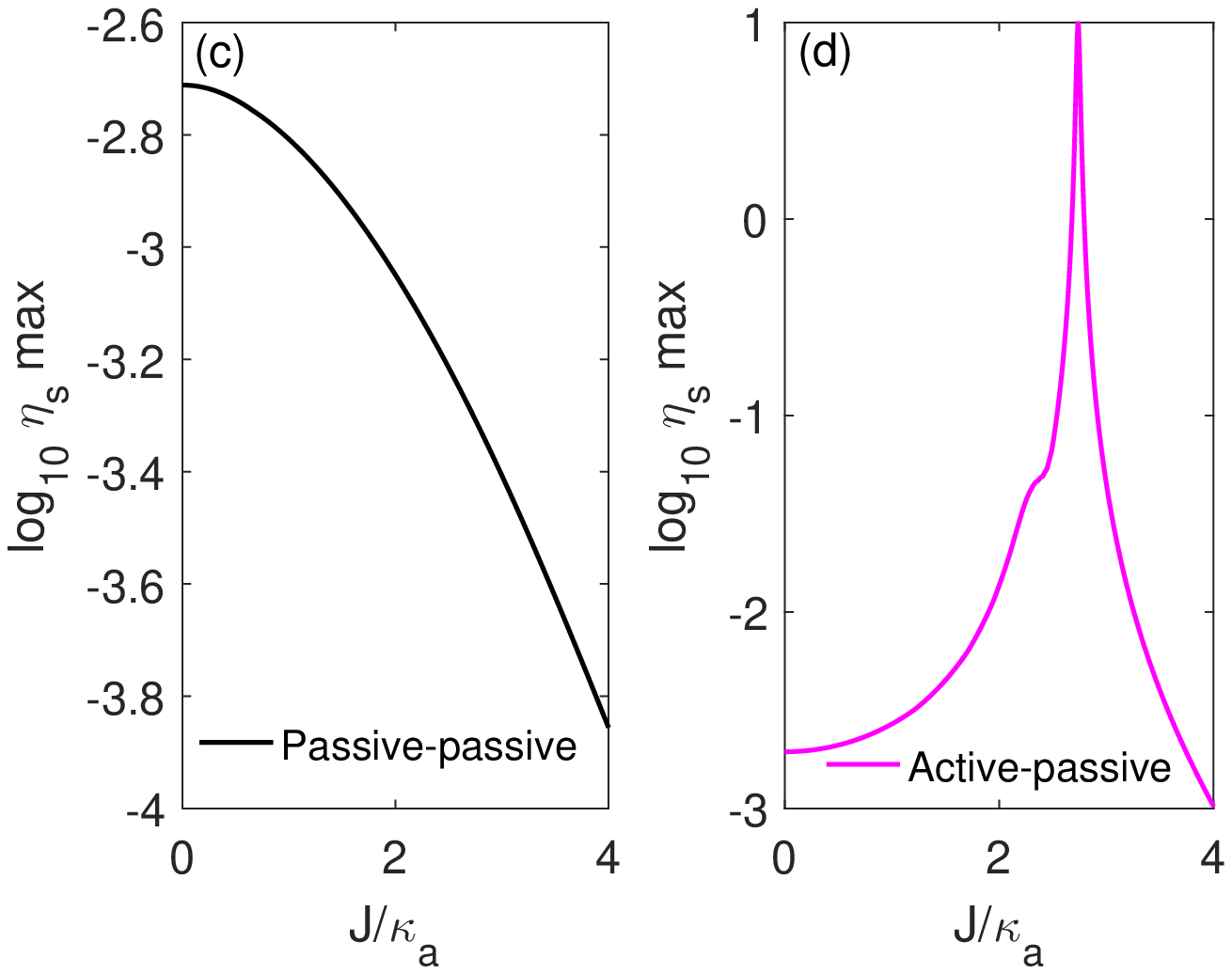}
\caption {The logarithm of the maximum value of ${\eta _s}$ varies with the coupling strength $J$ for (a) the passive-passive system and near $\Omega /{\omega _m} =  - 1$; (b) the active-passive system and near $\Omega /{\omega _m} =  - 1$; (c) the passive-passive system and near $\Omega /{\omega _m} = 1$; (d) the active-passive system and near $\Omega /{\omega _m} = 1$. The other parameter values are the same as in Fig.~\ref{fig3} except for ${P_1} = 210.3{\rm{ }}\mu {\rm{W}}$.}
\label{fig4}
\end{figure}

We further understand the dependence of the second-order sideband on the coupling strength $J$. The variation of the maximum values of ${\eta _s}$ with coupling strength $J/{\kappa _a}$ for the passive-passive and the PT-symmetric system are depicted in Fig.~\ref{fig4}. In the limit range of $J/{\kappa _a} \to 0$, the composite system can be simplified as an optomechanical system with atom-cavity-resonator coupling. The corresponding starting point of ${\eta _s}$ is the same. As illustrated in Fig.~\ref{fig4}(a), the second-order sideband generation of the passive-passive system near $\Omega  =  - {\omega _m}$ increases exponentially in the range $0 < J/{\kappa _a} < 4$. We show in Fig.~\ref{fig4}(b) the variation of the maximum value of ${\eta _s}$ becomes significantly different in a passive-active system. ${\eta _s}_{\max }$ near $\Omega  =  - {\omega _m}$ decreases with the increase of $J$ and this is consistent with the phenomenon shown by Fig.~\ref{fig3}(b). Especially in the vicinity of $J/{\kappa _a} = 2.8$, there is a sharp increase of the minimum value to a maximum value. In addition, we can see clearly from Fig.~\ref{fig4}(c) that a strong photon-tunneling strength has an adverse effect on the generation of the second-order sideband in the passive-passive system. ${\eta _s}_{\max }$ near $\Omega  = {\omega _m}$ decreases exponentially with the increase of $J$. However, the result of the PT-symmetric system is different from that of the passive-passive optomechanical system. As shown in Fig.~\ref{fig4}(d), ${\eta _s}_{\max }$ experiences exponential growth with the increase of $J$, and reaches its maximum value in the vicinity of $J/{\kappa _a} = 2.8$; then, it goes through sharp reduction, which means that the system transits from broken-PT-symmetric phase to PT-symmetric phase. Compared with the passive-passive system, the values of ${\eta _s}_{\max }$ in the passive-active system are overall improved [see Figs.~\ref{fig4}(a)-\ref{fig4}(c)]. This implies that PT-symmetric structure can greatly improve the efficiency of the second-order sideband by changing photon-tunneling strength, especially in the vicinity of EP. The efficiency of the OSSG is greatly improved in the PT-symmetric arrangement due to strong field localization-induced dynamical intensity accumulation, which significantly enhances nonlinearity in the PT-symmetric system~\cite{liu2016metrology,zhang2015giant}.

\subsection{THE OUTPUT EFFECTS OF ATOMIC ENSEMBLE ON THE GENERATION OF FIRST-ORDER AND SECOND-ORDER SIDEBANDS}

\begin{figure}[htbp]
\centering\includegraphics[width=14cm]{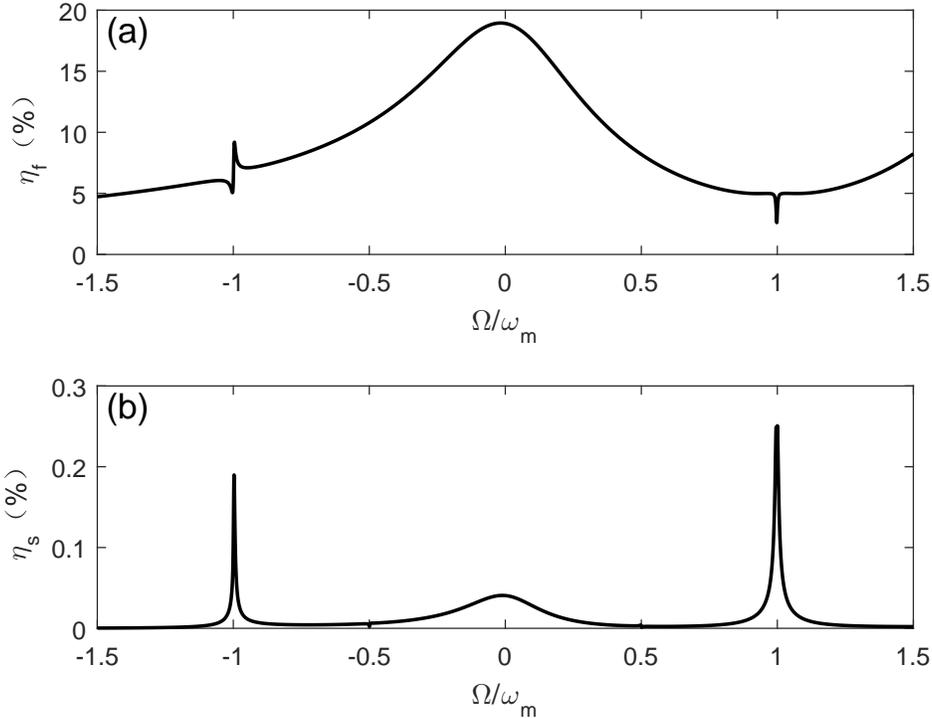}
\caption {(a) Calculation result of ${\eta _f}$ as a function of the detuning $\Omega /{\omega _m}$ for the PT-symmetric system. (b) Calculation result of ${\eta _s}$  as a function of the detuning $\Omega /{\omega _m}$ for the PT-symmetric system. $J = 0.55{\kappa _a}$, the other system parameters are the same as in Fig.~\ref{fig3}.}
\label{fig5}
\end{figure}

Through Eq. (\ref{19}), we can find that there are two different processes in the OSSG, i.e., a direct OSSG process $\left[ {{\theta _1}\left( \Omega  \right) \sim X_1^2} \right]$ produced by the two-phonon converted process of the pump field, and an indirect OSSG process $\left\{ {\left[ {{\theta _4}\left( \Omega  \right) - {a_s}{\theta _3}\left( \Omega  \right)} \right] \sim A_1^ - {X_1}} \right\}$, which is a single-phonon converted process of the upconverted first-order sideband. As we all know, it is more difficult to induce the conversion process of two-phonon than that of one-phonon, thus the OSSG mainly depends on indirect process. In other words, the first-order sideband process will directly affect the OSSG. Then we draw the first and second order sideband efficiencies respectively. As shown in Fig.~\ref{fig5}, the OFSG and OSSG have obvious changes in the vicinity of $\Omega /{\omega _m} =  - 1$ or $\Omega /{\omega _m} =   1$. Compared with the generation efficiency of the first-order sideband, the second-order sideband efficiency is significantly weaker. The OFSG and OSSG have a local minimum on resonance $\Omega /{\omega _m} = 1$. This can be intuitively explained as follows. In the optomechanical system, the efficiency of the first and second order sidebands depend on the power of the pump light and the optomechanical coupling strength. Their efficiency is usually very small. Due to in the process of the OSSG, the conversion process of the two-phonon into the second-order sideband is very weak, and its generation mainly depends on the upconverted first-order sideband. Under the resonance condition of $\Omega /{\omega _m} = 1$, the anti-Stokes field is resonantly enhanced. Moreover, the upconverted first-order sideband is suppressed due to the destructive interference between the probe field and the anti-Stokes field generated in the cavity, thus leading to the suppression of the second-order sideband. However, when the resonance condition is no longer satisfied $\left( {\Omega /{\omega _m} =  - 1} \right)$, the anti-Stokes field is also no longer enhanced by resonance. The efficiency of the second-order sideband is also enhanced under this premise. The existence of atoms introduces new nonlinearity, thus the second-order sideband can be generated not only from the nonlinearity of the optomechanical interaction, but also from the nonlinearity of atomic ensemble. As shown in Fig.~\ref{fig5}, the OSSG has a significant increase near $\Omega /{\omega _m} =  - 1$.

\begin{figure}[htbp]
\centering\includegraphics[width=14cm]{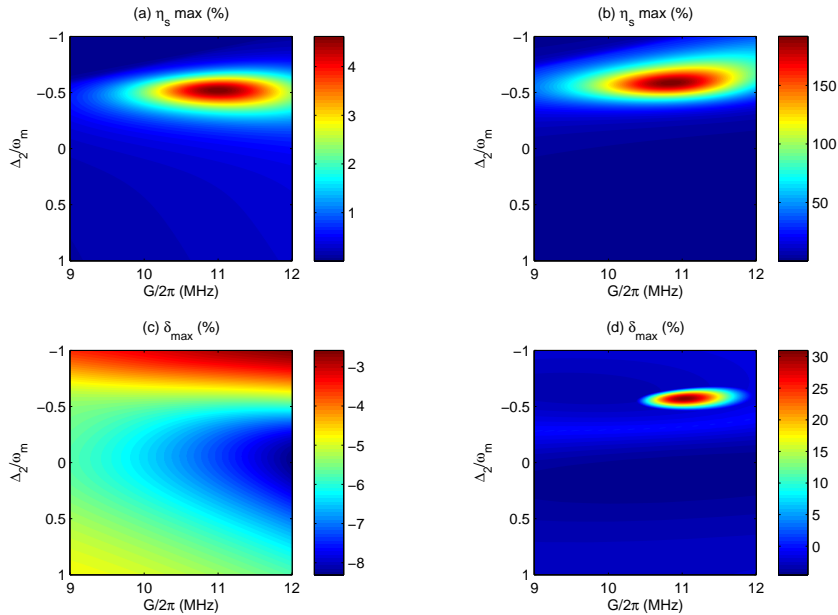}
\caption {Maximum efficiency ${\eta _s}$ of the OSSG around (a) $\Omega /{\omega _m} =  - 1$ and (b) $\Omega /{\omega _m} = 1$ as a function of the atom-pump detuning ${\Delta _2}$ and the atom-cavity coupling strength $G$.
The maximum value ${\delta _{\max }}$ of the difference between the efficiency of the OSSG and that of the OFSG around (c) $\Omega /{\omega _m} =  - 1$ and (d) $\Omega /{\omega _m} = 1$ as a function of the atom-pump detuning ${\Delta _2}$ and the atom-cavity coupling strength $G$. The other parameters are the same as in Fig.~\ref{fig5}.}
\label{fig6}
\end{figure}

The nonlinearity of the physical medium of the system plays an important role in the generation of second-order or high-order sidebands. Next, we further study the influence of atomic ensemble on the generation of first-order and second-order sidebands. We are considering whether the efficiency of the OSSG can exceed that of the OFSG. In view of this, we draw maximum efficiency ${\eta _s}$ of the OSSG around $\Omega /{\omega _m} =  \pm 1$ versus the atom-pump detuning ${\Delta _2}$ and the atom-cavity coupling strength $G$ in Figs.~\ref{fig6}(a)-\ref{fig6}(b). Moreover, the difference between the efficiency of the OSSG and that of the OFSG can be described by
\begin{eqnarray}\label{24}
\delta  = {\eta _s} - {\eta _f}.
\end{eqnarray}

In Figs.~\ref{fig6}(c)-\ref{fig6}(d), we plot the maximum value of the difference ${\delta _{\max }}$ near the blue probe-pump detuning resonant case and red one. We find that the efficiency of the OSSG is greatly improved for $G \in \left( {9.5,12} \right){\rm{MHz}}$ and ${\Delta _2} \sim  - 0.5{\omega _m}$ [see Figs.~\ref{fig6}(a)-\ref{fig6}(b)]. The nonlinearity-induced enhancement of the first-order or second-order sidebands near $\Omega /{\omega _m} =  \pm 1$ by the atom-cavity coupling strength seems mainly concentrated on ${\Delta _2} \sim  - 0.5{\omega _m}$. Especially, as we can see from Fig.~\ref{fig6}(b) the efficiency ${\eta _s}$ near $\Omega /{\omega _m} = 1$ can reach to about $150\% $ in the regions ${\Delta _2} \sim  - 0.5{\omega _m}$ and $G = 2\pi  \times 11{\rm{ MHz}}$. This means that the amplitude of the output second-order sideband is $1.5$ of the amplitude of the input. Furthermore, ${\delta _{\max }}$ are negative and we note that the blue area is mainly concentrated in the regions ${\Delta _2} \in \left( { - 0.5,0.5} \right){\omega _m}$ [see Fig.~\ref{fig6}(c)]. It is obvious that the efficiency of the OSSG is still less than the efficiency of the OFSG in the vicinity of $\Omega /{\omega _m} =  - 1$. However, the efficiency of the OSSG is much higher (above $30\% $) than that of the OFSG in the regions $G \in \left( {10.5,12} \right){\rm{MHz}}$ and ${\Delta _2} \sim  - 0.5{\omega _m}$ for the blue probe-pump detuning resonant case ($\Omega /{\omega _m} = 1$). This implies that the atom-pump detuning ${\Delta _2}$ has more significant effect on the efficiency enhancement of the OSSG in the case of the resonance condition. Through the above analyses, we show that the existence of atoms makes the first and second order sidebands of the system more flexible and adjustable. Moreover, it is proved that the efficiency of the OSSG is higher than the efficiency of the OFSG there is a region formed by the atom-pump detuning and the atom-cavity coupling strength.

\subsection{ENHANCED OUTPUT EFFECTS OF THE PUMP POWER ON THE FIRST- AND SECOND-ORDER SIDEBANDS}

\begin{figure}[htbp]
\centering\includegraphics[width=7cm]{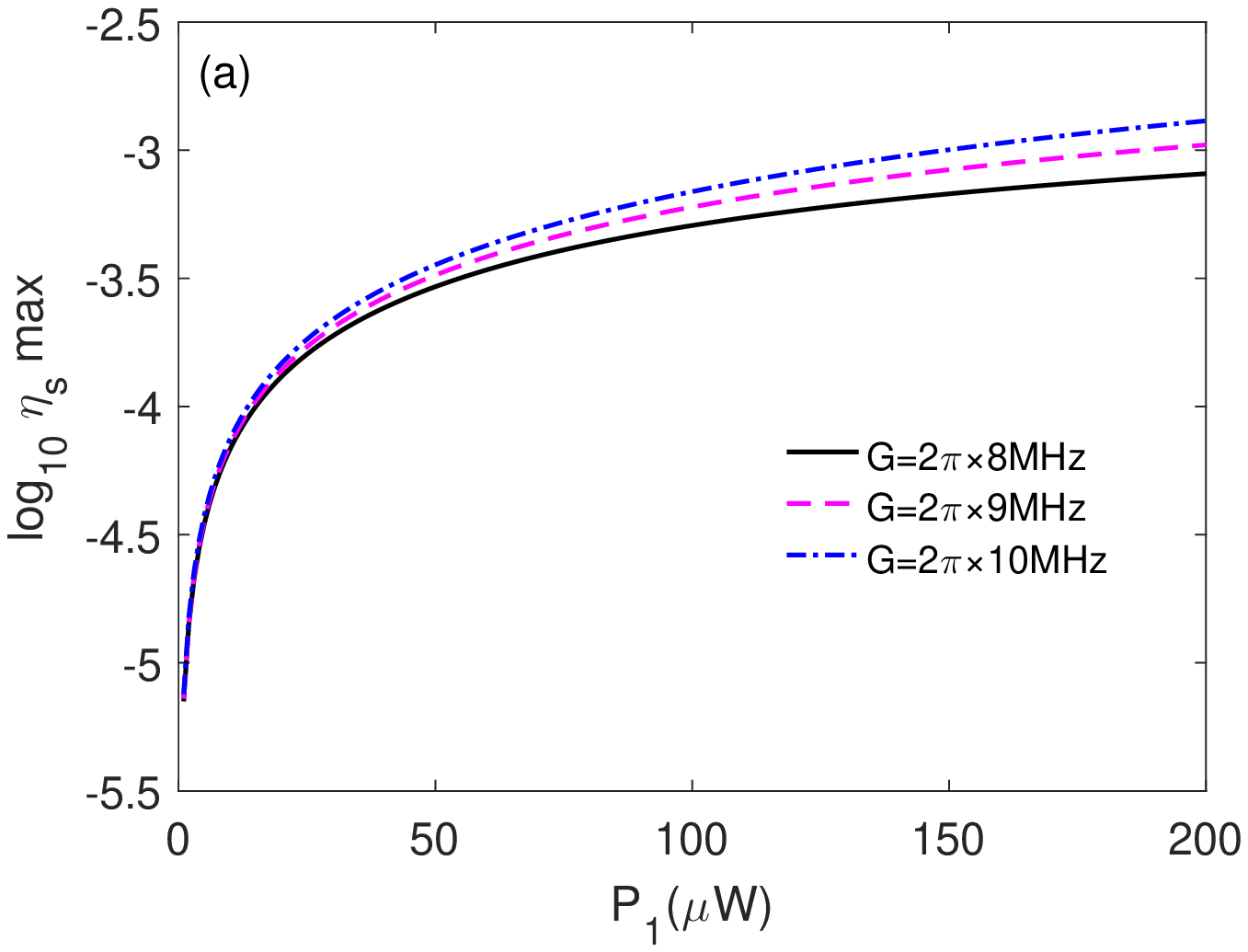}
\includegraphics[width=7cm]{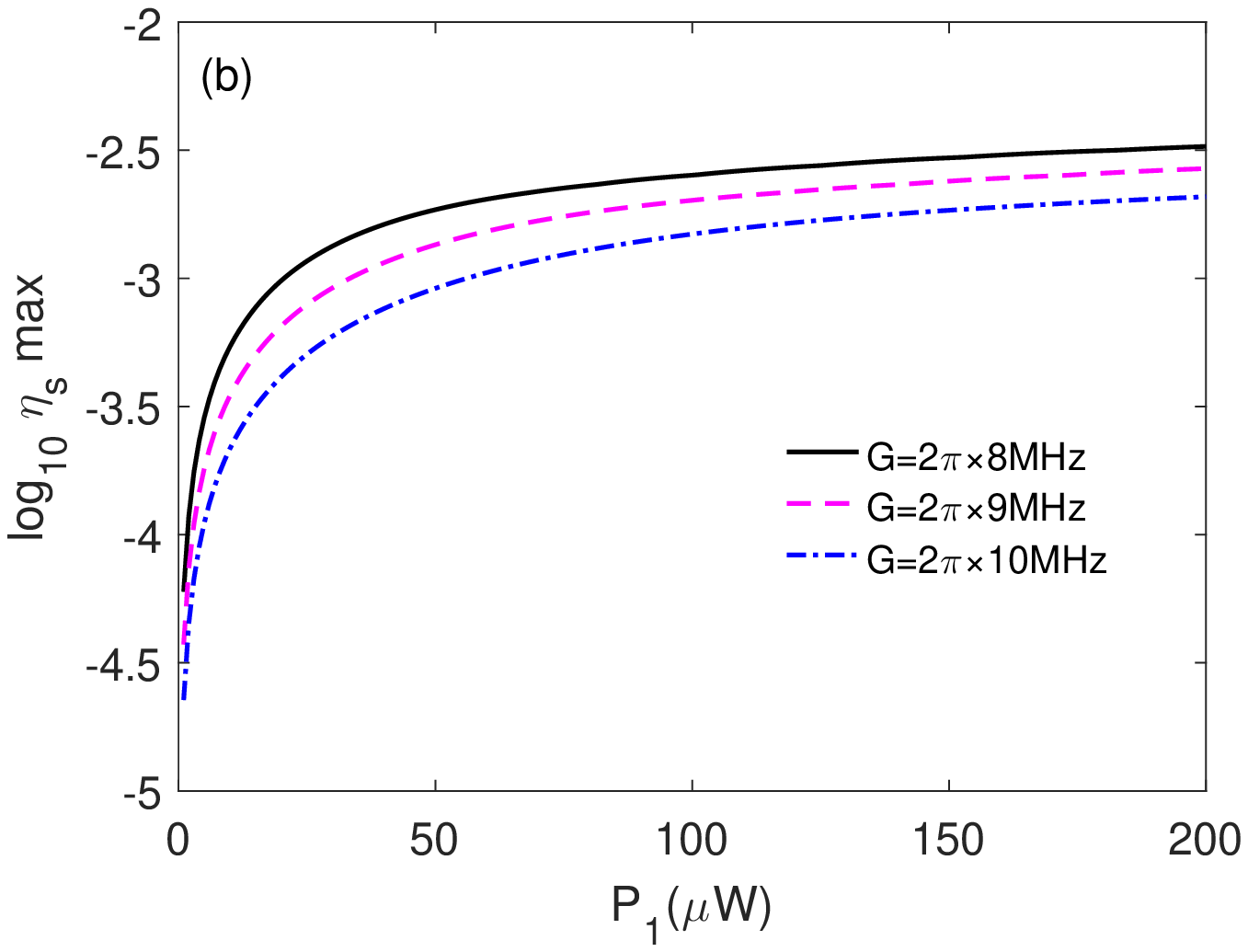}
\caption {The simulation results of maximum efficiency ${\eta _s}$ of the OSSG vary with the pump power ${P_1}$ around (a) $\Omega /{\omega _m} = -1$ and (b) $\Omega /{\omega _m} = 1$. The solid, dashed, and dash-dotted lines, respectively, correspond to the different the atom-cavity coupling strengths $G = 2\pi  \times 8{\rm{ MHz}}$, $G = 2\pi  \times 9{\rm{MHz}}$ and $G = 2\pi  \times 10{\rm{MHz}}$. The other parameters are the same as in Fig.~\ref{fig5}.}
\label{fig7}
\end{figure}

It is well known that the pump power ${P_1}$ will inevitably affect the OSSG. In order to explicitly show the influence of the pump field on the OSSG, we draw the peak value versus the power ${P_1}$ in Fig.~\ref{fig7}.

According to Fig.~\ref{fig7}, the enhancement of second-order sideband generation is remarkable when the power of the pump field is small, and the efficiency increases as the pump power increases. With the further increase of the pump power to a higher value, ${\eta _s}$ goes to be stabilized. The solid, dashed, and dash-dotted lines correspond to $G = 2\pi  \times 8{\rm{ MHz}}$, $G = 2\pi  \times 9{\rm{MHz}}$ and $G = 2\pi  \times 10{\rm{MHz}}$ in Fig.~\ref{fig7}, respectively. The efficiency of the OSSG near $\Omega /{\omega _m} =  - 1$ can be increased by increasing the atom-cavity coupling strength [see Fig.~\ref{fig7}(a)]. The existence of atoms enhances the nonlinearity of the system. On the contrary, with the increase of the atom-cavity coupling strength, the efficiency of the OSSG near $\Omega /{\omega _m} = 1$ reduces, as is shown in Fig.~\ref{fig7}(b). The physical interpretation of this phenomenon is that the first-order sideband is suppressed when resonance due to the destructive interference between the probe field and the anti-Stokes field produced in the cavity.


\section{Conclusions}
This paper provides an in-depth investigation of the OSSG in a PT-symmetric optomechanical system, in which one active cavity with a gain is directly coupled to a passive cavity. The passive cavity traps the atomic ensemble. We first consider the effect of the oscillating frequency of the mechanical resonator on the transmission in a PT-symmetric system without the atomic ensemble. It was clear that the transmission intensity can be improved by increasing the value of the oscillating frequency. And there are linear and nonlinear regions in the relationship between them. In the vicinity of EP, the efficiency of the OSSG enhances sharply not only for the blue probe-pump detuning resonant case but also for the red one in our system. Moreover, the influence of the atomic ensemble on the efficiency of the OSSG has been studied. It is proved that a large efficiency of the OSSG can be obtained. The efficiency will be significantly higher than the efficiency of the OFSG by modulating the atom-pump detuning and the atom-cavity coupling strength together. In addition, we also focused our attention to the effect of the pump power variations and showed that the efficiency of the OSSG is robust when the pump power is large enough. The efficiency of the OSSG can be increased by increasing the pump power or the atom-cavity coupling strength. These results indicate the possibility to enhance the nonlinear optical properties or steer complex integrated PT devices.

\section{Acknowledgments}
The authors are grateful to all of those who were involved in this work. This project was supported by National Natural Science Foundation of China (NSFC) (Grant Nos. 61368002, 91736106, 11674390, and 91836302), Open Research Fund Program of the State Key Laboratory of Low-Dimensional Quantum Physics (KF201711), the Foundation for Distinguished Young Scientists of Jiangxi Province (20162BCB23009) and the Graduate Innovation Special Fund of Jiangxi Province (YC2019-S102).

\section{Appendix}
By substituting Eqs. (\ref{11})-(\ref{17}) into the quantum fluctuation of Eqs. (\ref{2})-(\ref{6}), we will obtain some equations about the coefficient of the first-order $A_1^ - $ and second-order $A_2^ - $ upper sidebands.
\begin{eqnarray}\label{25}
\left[ {i\left( {\bar \Delta  + \Omega } \right) - \frac{{{\kappa _a}}}{2}} \right]A_1^ -  = iJB_1^ -  + iGC_1^ -  + i{g_1}\left( {{a_s}{X_1} + A_1^ + {X_2} + A_2^ - X_1^ * } \right) - \sqrt {\eta {\kappa _a}} {\varepsilon _p},
\end{eqnarray}
\begin{eqnarray}\label{26}
\left[ {i\left( {\bar \Delta  - \Omega } \right) - \frac{{{\kappa _a}}}{2}} \right]A_1^ +  = iJB_1^ +  + iGC_1^ +  + i{g_1}\left( {{a_s}X_1^ *  + A_1^ - X_2^ *  + A_2^ + {X_1}} \right),
\end{eqnarray}
\begin{eqnarray}\label{27}
\left( {\omega _m^2 - {\Omega ^2} - i\Omega {\Gamma _m}} \right)m{X_1} =  - \hbar {g_1}\left( {{a_s}A{{_1^ + }^ * } + {a_s}A_1^ -  + A{{_1^ - }^ * }A_2^ -  + A_1^ + A{{_2^ + }^ * }} \right),
\end{eqnarray}
\begin{eqnarray}\label{28}
\left[ {i\left( {\bar \Delta  + 2\Omega } \right) - \frac{{{\kappa _a}}}{2}} \right]A_2^ -  = iJB_2^ -  + iGC_2^ -  + i{g_1}\left( {{a_s}{X_2} + A_1^ - {X_1}} \right),
\end{eqnarray}
\begin{eqnarray}\label{29}
\left[ {i\left( {\bar \Delta  - 2\Omega } \right) - \frac{{{\kappa _a}}}{2}} \right]A_2^ +  = iJB_2^ +  + iGC_2^ +  + i{g_1}\left( {{a_s}X_2^ *  + A_1^ + X_1^*} \right),
\end{eqnarray}
\begin{eqnarray}\label{30}
\left( {\omega _m^2 - 4{\Omega ^2} - 2i\Omega {\Gamma _m}} \right)m{X_2} =  - \hbar {g_1}\left( {{a_s}A{{_2^ + }^ * } + a_s^*A_2^ -  + A_1^ - A{{_1^ + }^ * }} \right).
\end{eqnarray}
By solving the above equations, we can work out the results of $A_1^ - $ and $A_2^ - $ as shown in Eq. (\ref{18}) and Eq. (\ref{19}).

\end{document}